\documentclass[11pt,a4]{article}
\usepackage{graphics}
\usepackage{comment}
\usepackage[numbers]{natbib}
\usepackage{amsmath}
\usepackage{multirow}
\usepackage{epstopdf}
\usepackage[a4paper,top=1.0cm,bottom=1.2cm,left=2cm,right=2cm,marginparwidth=1.5cm]{geometry}

\title { \bf Dynamical Complexity of Short and Noisy Time Series}
\author{Nithin Nagaraj$^\dagger$ and Karthi Balasubramanian$^*$}

\date{\small $^\dagger$Consciousness Studies Programme, National Institute of Advanced Studies\\ Indian Institute of Science Campus, Bengaluru 560 012, India. \\{\bf \small Email:~nithin@nias.iisc.ernet.in} \\
$^*$Department of Electronics \& Communication Engg., Amrita School of Engg., Coimbatore \\ Amrita Vishwa Vidyapeetham, Amrita University, India. \\{\bf \small Email:~b\_karthi@cb.amrita.edu}\\ November 25, 2016}

\begin{document}
\maketitle

\abstract{
Shannon Entropy has been extensively used for characterizing complexity of time series arising from chaotic dynamical systems and stochastic processes such as Markov chains. However, for short and noisy time series, Shannon entropy performs poorly. Complexity measures which are based on lossless compression algorithms are a good substitute in such scenarios. We evaluate the performance of two such {\it Compression-Complexity Measures} namely Lempel-Ziv complexity ($LZ$) and Effort-To-Compress ($ETC$) on short time series from chaotic dynamical systems in the presence of noise. Both $LZ$ and $ETC$ outperform Shannon entropy ($H$) in accurately characterizing the dynamical complexity of such systems. For very short binary sequences (which arise in neuroscience applications), $ETC$ has higher number of distinct complexity values than $LZ$ and $H$, thus enabling a finer resolution. For two-state ergodic Markov chains, we empirically show that $ETC$ converges to a steady state value faster than $LZ$. {\it Compression-Complexity Measures} are promising for applications which involve short and noisy time series.
} 
\maketitle
\section{Introduction}
\label{intro}
Claude Shannon introduced the idea of `Entropy' as a quantitative measure of information in 1948~\cite{Shannon1948} when he was building a mathematical theory of communication. The notion of entropy had already been proposed in thermodynamics (Clausius, 1965) and in statistical physics (Boltzmann and Gibbs, 1900s). Shannon entropy of a discrete random variable is defined as:
\begin{equation}
    H(\chi) = - \sum_{i=1}^{M} p_i \log_2(p_i)~~~\mbox{bits/symbol}, \label{eq:eqnENT}
\end{equation}
where $\chi$ is the random variable with $M$ possible events and the probability of occurrence of the $i$-th event is given by $p_i>0$. The maximum value of the concave function $H(prob.)$ is achieved for a uniform random variable with all events equally likely ($H = \log_2(M)$ bits).

Apart from playing a fundamental role in communications, information and coding theory, Shannon entropy is also used to characterize the complexity of a time series. Low entropy of a time series  indicates low complexity (less randomness and hence more structure) whereas a higher value of entropy of a time series would imply a higher complexity (more randomness and hence less structure). This is because, Shannon entropy characterizes the degree of compressibility of an input sequence. Today, Shannon entropy (or $H$), and some of its related information theoretic measures (such as mutual information, conditional entropy etc.), continue to be widely used as measures of dynamical complexity in several applications. It is used in biomedical applications~\cite{Entropy1}, for eg., as a pattern classification tool in heart rate variability analysis~\cite{Entropy2}; to measure structural and dynamical complexity of networks~\cite{Entropy3} and communication complexity~\cite{Entropy4}; for biological sequence analysis in bioinformatics~\cite{Entropy5, Entropy6}; in econometric/financial time series analysis~\cite{Entropy7, Entropy8, Entropy9}; and not to miss out on the various entropic forms in physics~\cite{Entropy10}. This is by no means an exhaustive list, but only serves as indicative of the diverse domains in which Shannon entropy is applied.

However, Shannon entropy ($H$) has serious drawbacks when the time series under consideration is short and noisy. In this work, we point out these limitations and propose the use of {\it Compression-Complexity} measures to overcome these limitations of Shannon entropy for characterizing dynamical complexity of short and noisy time series. {\it Compression-Complexity} measures shall be defined as complexity measures based on lossless compression algorithms. This is the subject matter discussed in sections 2 and 3 of this paper. 

Signals that are seen in real world are never completely random in nature, though they may be stochastic in origin. In several instances,  these signals behave as information sources that may be modelled as Markov or hidden Markov processes. Markov chains, named after Andrei Andreievich Markov (1856-1922), is a type of random process which has the property that the current state of the system depends only on its immediate past state\footnote{This is the definition of a 1-order Markov process.} and not on the sequence of past states prior to that.  The transition from one state to another state is captured by transition probabilities. Markov chains have played a vital role for modeling in statistical mechanics. Dating back to the urn models for mixing of D. Bernoulli (1769), Laplace (1812) and Ehrenfest (1907), these are simple examples of Markov chain models (known as random walks).

Many real world systems behave like Markov sources that produce signals that may be recreated using finite chain Markov process models. \textit{E.g.}, the patterned structure of heart-beat intervals  \cite{heartbeat_markov1, heartbeat_markov2,heartbeat_markov3,
heartbeat_markov4}, base compositions of DNA sequences \cite{dna_markov1,dna_markov2, dna_markov3,dna_markov4}, decomposition and recognition of speech \cite{speech_markov1,speech_markov2, speech_markov2,speech_markov4}, language scripts modelling \cite{script_markov1,script_markov2, script_markov3}, information sources in communication systems \cite{information_markov1,information_markov2, information_markov3}, trend prediction of stock indices~\cite{fin_markov1} and analysis of share prices~\cite{fin_markov2}, can all be mathematically viewed as Markov processes/chains. Hence, a study of the performance of complexity measures on data produced from Markov chains would be a good indication of its
performance on real world signals. In section 4, we simulate a 2-state Markov chain and evaluate the performance of {\it Compression-Complexity} in characterizing its complexity.

We conclude with future research directions in the last section.

\begin{figure}[!h]
\centering
\resizebox{1.0\columnwidth}{!}
{\includegraphics{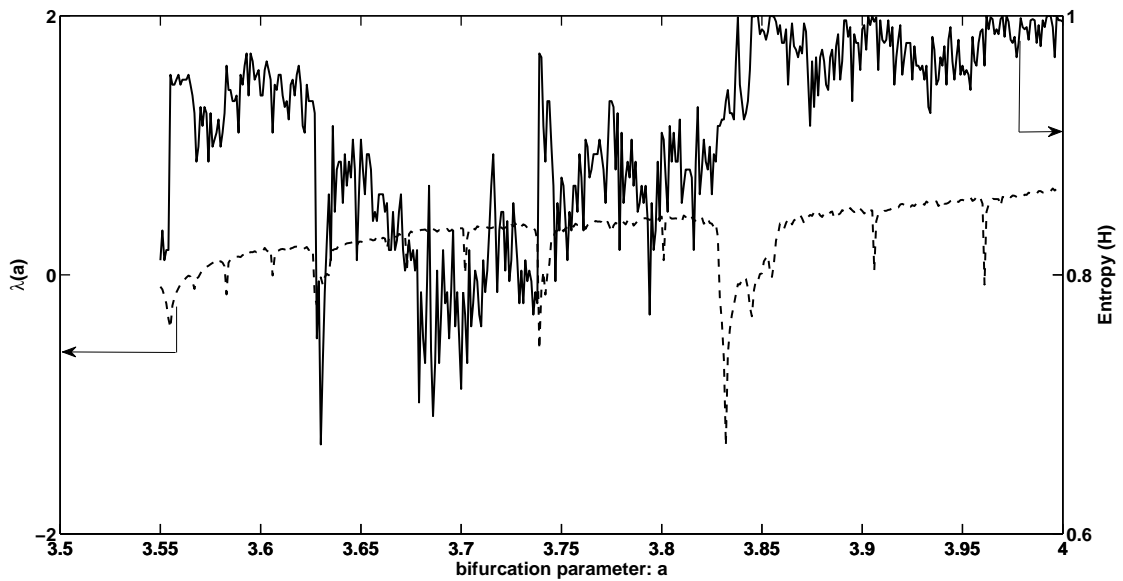}}
\caption{Shannon Entropy $H$ (solid) and Lyapunov exponent $\lambda(a)$ (dotted) for the logistic map as the bifurcation parameter $a$ is varied from $3.55$ to $4.0$ (in steps of $0.01$). It can be seen that $H$ is poorly correlated with $\lambda(a)$. This is also indicated quantitatively by a low value of Pearson's correlation coefficient $=0.2721$ between $H$ and $\lambda(a)$. Here, we have chosen $L=200$ and number of bins $= 4$ for the symbolic sequence.}
\label{figure:figH}
\end{figure}

\section{Limitations of Shannon entropy as a measure of dynamical complexity}
\label{sec:2}
In order to explicitly demonstrate the limitations of Shannon entropy as a measure of dynamical complexity, we consider the chaotic dynamical system which is known as the Logistic map~\cite{Alligood}.  The governing equation of this dynamical system is:\\
\begin{equation}
    x_{n+1} = a x_{n}(1-x_{n}), \label{eq:eqnLogi}
\end{equation}
where $x_n$ is the value at discrete time step $n > 0$, $x_1$ is chosen randomly from $(0, 1)$ and $a$ is the bifurcation parameter ($0 \leq a \leq 4$). The Logistic map is known to exhibit chaos for certain values of the bifurcation parameter `$a$'~\cite{Alligood}. By varying the bifurcation parameter $a$, one can obtain time series $\{ x_n \}$ which exhibits periodic behaviour (for eg., $a=3.83$), weak chaos ($a=3.75$), strong chaos ($a=3.9$) or even complete chaos ($a=4.0$ where there are no attracting periodic orbits). It is intuitive that periodic time series is of low complexity whereas weak chaos has higher complexity and strong chaos is of even higher complexity. An even higher complexity would be manifested by complete chaos, though we would expect a uniform random sequence from a stochastic source to be of the highest complexity. Thus, the time series obtained by the Logistic map serves as a test data-set which can be used to determine whether Shannon entropy can quantitatively characterize the complexity correctly and also the order of complexities (from weak chaos to strong chaos to complete chaos).  We shall compare this with Lyapunov exponent $\lambda$, which measures the degree of sensitive dependence to initial conditions.  The Lyapunov exponent $\lambda$ is zero for a periodic time series and positive for chaotic time series (and increases in value with increasing strength of chaos in the system, and is very high for strong chaos). $\lambda$ serves as an excellent quantitative measure of complexity for dynamical systems whose equations are known, because, in such instances it can be computed using an analytical expression.  For the Logistic map, $\lambda$ is given by the following expression:
\begin{equation}
    \lambda(a) = \lim_{L \rightarrow \infty} \frac{1}{L} \sum_{i=1}^{L} \ln(|a(1-2x_i)|), \label{eq:eqnLogiLE}
\end{equation}
where $L$ stands for length of time series and $\{ x_i \}$ is the time series generated by Equation~\ref{eq:eqnLogi} starting from a randomly chosen initial value $x_1$ in $(0, 1)$.  For simulation purposes, we take $L$ as the actual length of the time series in Equation~\ref{eq:eqnLogiLE} (and drop the `limit'). 

In order to test the performance of Shannon entropy $H$ to characterize dynamical complexity of the logistic map, we vary the bifurcation parameter $a$ from $3.5$ to $4.0$ and generate time series of length $L=200$ from randomly chosen initial value $x_1$ in $(0, 1)$. The Lyapunov exponent $\lambda(a)$ for each time series is computed using Equation~\ref{eq:eqnLogiLE}.  Also, we compute the Shannon entropy $H$ of the {\it symbolic sequence} of each time series using Equation~\ref{eq:eqnENT}. A symbolic sequence is a sequence of symbols for each corresponding value of the time series. We have chosen four equal sized bins spanning the entire range of the time series $[Min, Max]$ and associate a unique symbol to each bin. Thus, the four bins would be $[Min, V_1]$, $[V_1, Mean]$, $[Mean, V_2]$ and $[V_2, Max]$ with the corresponding symbols being $A$, $B$, $C$ and $D$ where $V_1 = \frac{Min+Mean}{2}$ and  $V_2 = \frac{Mean+Max}{2}$. Figure~\ref{figure:figH} depicts the graph of entropy $H$ and the Lyapunov exponent $\lambda(a)$. It can be seen that $H$ is poorly correlated with $\lambda(a)$. This is also indicated quantitatively by a low value of Pearson's correlation coefficient $=0.2721$ between $H$ and $\lambda(a)$. 

\begin{figure}[!h]
\centering
\resizebox{1.1\columnwidth}{!}
{
\includegraphics{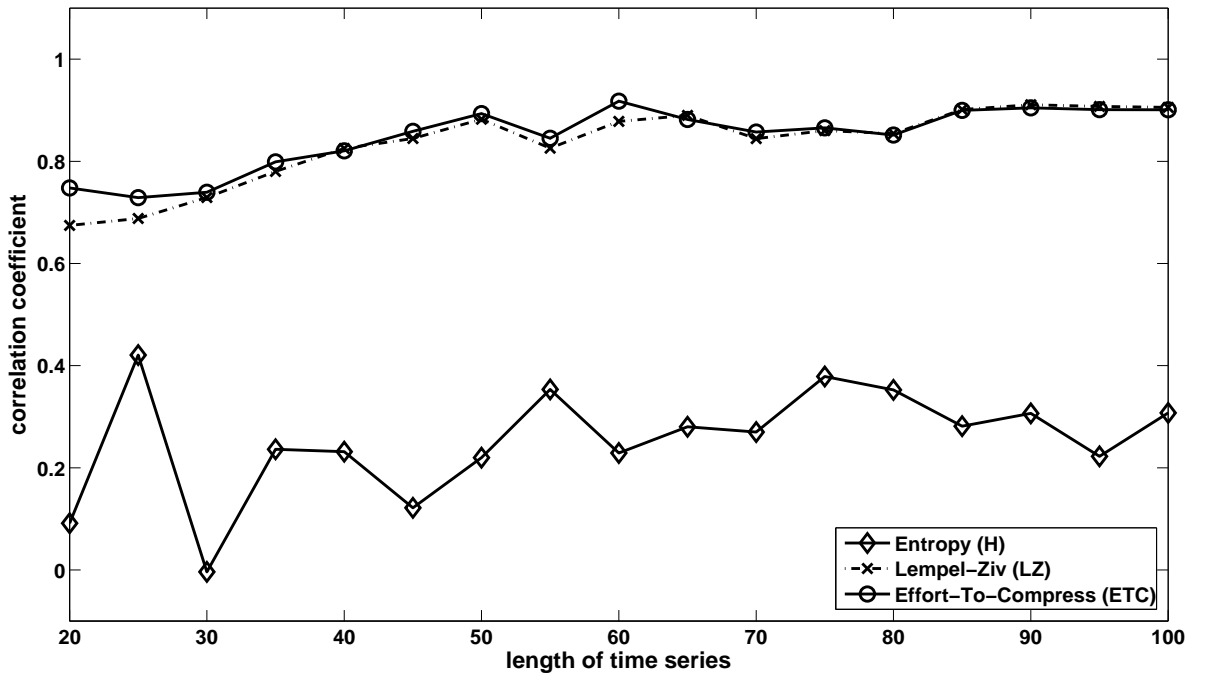}}
\caption{The correlation of Shannon entropy ($H$), Lempel-Ziv complexity ($LZ$) and Effort-To-Compress complexity ($ETC$) with Lyapunov exponent ($\lambda$) for the logistic map ($a=3.55$ to $4.0$) for different lengths of time series. Entropy is very poorly correlated with Lyapunov exponent, whereas both $LZ$ and $ETC$ show very good correlation. We have used 4 symbols in all cases.}
\label{figure:comparison}
\end{figure}

Thus, the empirically determined Shannon entropy $H$ is a poor indicator of dynamical complexity, especially for short time series ($L=200$ and lesser). The performance of $H$ as the length of time series is varied is depicted in Figure~\ref{figure:comparison}.

\section{Compression-Complexity Measures}
\label{sec:3}
As noted in the previous section, there are serious limitations to Shannon entropy as a measure of dynamical complexity for short time series arising from chaotic dynamical systems. This motivates us to find alternate measures to characterize the dynamical complexity in such scenarios. A class of measures which come to our rescue is what we call as {\it Compression-Complexity Measures}, or CCM for short.

Compression-Complexity Measures (CCM) are those measures of complexity which are derived from lossless data compression algorithms. It is well acknowledged that data compression algorithms are not only useful for compression of data for efficient transmission and storage, but also act as models for learning and statistical inference~\cite{Cilibrasi}. One example of such a measure is the popular Lempel-Ziv Complexity (LZ)~\cite{LZComplexity} which is closely related to the universal dictionary-based lossless compression algorithm by the same authors~\cite{LZCompression}. LZ complexity measure has been widely used in a number of fields such as - in biomedical applications~\cite{LZ1, LZ2}, for estimating entropy of spike trains~\cite{LZ3}, data analysis for random boolean networks~\cite{LZ4}, in studying transition between stationary and non-stationary chaos in a one-dimensional non-hyperbolic chaos map~\cite{LZ5}, in designing a new distance measure for phylogenetic tree construction~\cite{LZ6}, for measuring complexity of genetic sequences~\cite{LZ7}, as well as in a number of financial time series analysis applications~\cite{LZ8, LZ9}.

Another example of a CCM is the Effort-To-Compress (ETC) complexity measure which was introduced in~\cite{ETC}. Unlike LZ complexity, ETC measures not the degree of compressibility, but rather the effort to compress the input sequence by means of a lossless compression algorithm. In particular, ETC is implemented using a lossless compression algorithm known as Non-Sequential Recursive Pair Substitution algorithm (NSRPS)~\cite{NSRPS}. ETC has been particularly useful for characterizing complexity of short and noisy time series~\cite{ETC} and has recently been applied for characterizing complexity of cardiovascular dynamics~\cite{PeerJ}.

In what follows, we shall briefly describe LZ and ETC complexity measures and how they are applied on an input sequence. Subsequently, we shall evaluate their performance on time series obtained from various chaotic dynamical systems and compare and contrast with performance of Shannon entropy for characterizing dynamical complexity.

\subsection{Lempel-Ziv Complexity (LZ)}
\label{sec:3.1} 
In order to compute the Lempel-Ziv complexity (or LZ) of an input time series $\{ x_i \}$, it has to be first converted to a {\it symbolic sequence} $\{ s_i \}$. Mathematically, this can be expressed as follows:
\begin{eqnarray*}
s_i & = & 0, \mbox{~~~if $Min$ $\leq x_i \leq$ $Mean$},\\
& = & 1,  \mbox{~~~if $Mean$ $< x_i \leq$ $Max$},
\end{eqnarray*}
\noindent where $1 \leq i \leq n$ (integer) for a time series $\{ x_i\}$ of length $n$. $Min$, $Mean$ and $Max$ are the minimum, mean and maximum of the entire time series respectively. Here, the symbolic sequence  $\{ s_i\}$ is of the same length as the input time series $\{ x_i\}$, but with only two symbols $0$ and $1$. Sometimes, we use the term `bins' to represent number of symbols in the symbolic sequence. In the above example, the symbolic sequence has two `bins'. This can be extend to $M$ bins by uniformly binning the entire range of the time series $[Min, Max]$ into $M$ equal sized bins and using corresponding symbols $\{ 0, 1, \ldots, M-1 \}$ (note: the symbols could well be $\{a, b, c, ..., m \}$).

The resulting symbolic sequence $S = \{ s_i\}_{i=1}^{i=n} = s_1 s_2 \ldots s_n$ is then parsed from left to right in order to identify the number of distinct patterns present. This method of parsing was proposed by Lempel and Ziv~\cite{LZComplexity} and this is closely related to the universal compression algorithm~\cite{LZCompression}. We reproduce below a very succinct description of the algorithm for computing LZ complexity, taken from Hu, Gao and Principe~\cite{LZ2}. Let $S = s_1 s_2 \cdots s_n$ denote a symbolic sequence; $S(i, j)$ denote a substring of $S$ that starts at position $i$ and ends at position $j$; $V(S)$ denote the set of all substrings $\{ S(i, j),  i = 1, 2, \cdots n ;  j \geq i \}$. For example, let $S = abc$, then $V(S) = {a, b, c, ab, bc, abc}$. The parsing mechanism involves a left-to-right scan of the symbolic sequence $S$. Start with $i = 1$ and $j =1$. A substring $S(i, j)$ is compared with all strings in $V$ $(S(i, j − 1))$ (Let $V(S(1,0)) = {}$, the empty set). If $S(i, j)$ is present in $V (S(1, j − 1))$, then increase $j$ by $1$ and repeat the process. If the substring is not present, then place a dot after $S(i, j)$ to indicate the end of a new component, set $i = j + 1$, increase $j$ by 1, and the process continues. This parsing procedure continues until $j = n$, where $n$ is the length of the symbolic sequence. For example, the sequence `$aacgacga$' is parsed as `$a.ac.g.acga.$'. By convention, a dot is placed after the last element of the symbolic sequence and the number of dots gives us the number of distinct words which is taken as the LZ complexity, denoted by $c(n)$. In this example, the number of distinct words (LZ complexity) is 4. In order to be able to compare the LZ complexity of sequences of different lengths, a normalized measure is proposed~\cite{LZ1}:
\begin{equation}
C_{LZ} = (c(n)/n)log_{\alpha}n,
\end{equation}
where $\alpha$ denotes the number of unique symbols in the input sequence. 
\begin{figure}[!h]
\centering
\resizebox{1.0\columnwidth}{!}
{\includegraphics{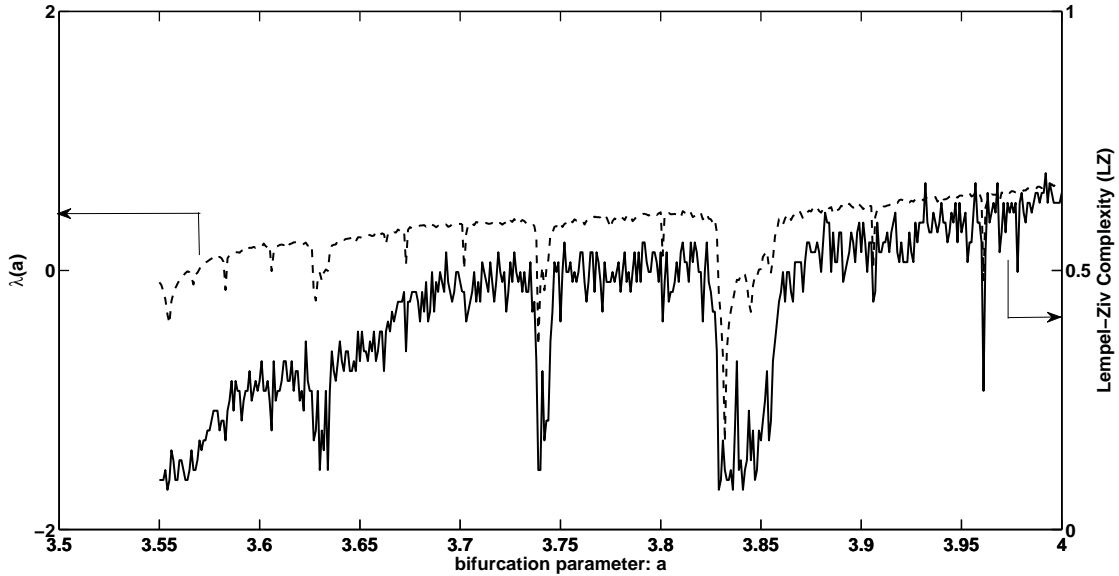}}
\caption{Lempel-Ziv complexity $LZ$ (solid) and Lyapunov exponent $\lambda(a)$ (dotted) for the logistic map as the bifurcation parameter $a$ is varied from $3.55$ to $4.0$ (in steps of $0.01$). The graphs are well correlated (positively). This is indicated by a value of Pearson's correlation coefficient $=0.8889$ between $LZ$ and $\lambda(a)$. Here, we have chosen $L=200$ and number of bins $= 4$ for the symbolic sequence.}
\label{figure:figLZ}
\end{figure}

Figure~\ref{figure:figLZ} depicts the graph of Lempel-Ziv complexity $LZ$ and the Lyapunov exponent $\lambda(a)$. The graphs are well correlated (positively). This is indicated by a high value of Pearson's correlation coefficient $=0.8889$ between $LZ$ and $\lambda(a)$. As the length of time series is varied, the correlation coefficient is consistently high as shown in Figure~\ref{figure:comparison}.

\subsection{Effort-To-Compress Complexity (ETC)}
\label{sec:3.2} 
Recently, we have proposed a new complexity measure known as Effort-To-Compress (ETC) which is based on the effort required by a lossless compression algorithm to compress a given sequence (Nagaraj, Balasubramanian, and Dey, 2013~\cite{ETC}). We have used a lossless compression algorithm known as Non-sequential Recursive Pair Substitution (NSRPS)~\cite{NSRPS}. The input time series is first converted into a {\it symbolic sequence} which was described in the previous subsection. The algorithm for compressing the resulting symbolic sequence proceeds as follows. At the first iteration, that pair of symbols which has the maximum number of occurrences is replaced by a new symbol. For example, the input sequence ‘11010010’ is transformed into ‘12202’ in the first iteration since the pair ‘10’ has maximum number of occurrences compared to all other pairs (‘00’, ‘01’ and ‘11’). In the second iteration, ‘12202’ is transformed to ‘3202’ (in fact all pairs are equally likely and we have chosen to replace `12'). The algorithm proceeds in this manner until the length of the transformed string shrinks to 1 or the transformed string becomes a constant sequence (at which stage the entropy of the transformed string is zero and the algorithm halts). In this example, the algorithm transforms the input sequence $‘11010010’ \mapsto ‘12202’ \mapsto ‘3202’ \mapsto ‘402’\mapsto ‘52’ \mapsto ‘6’$. 

The ETC complexity measure is defined as $N$, the number of iterations required for the input sequence to be transformed to a constant sequence through the usage of NSRPS algorithm. $N$ is always a non-negative integer that is bounded between 0 and $L-1$, where $L$ stands for the length of the input symbolic sequence. The normalized version of the measure is given by: $\frac{N}{(L-1)}$. Note that $0 \leq \frac{N}{(L-1)}  \leq 1$. Please refer to~\cite{ETC} for further details.

\begin{figure}[!h]
\centering
\resizebox{1.0\columnwidth}{!}
{\includegraphics{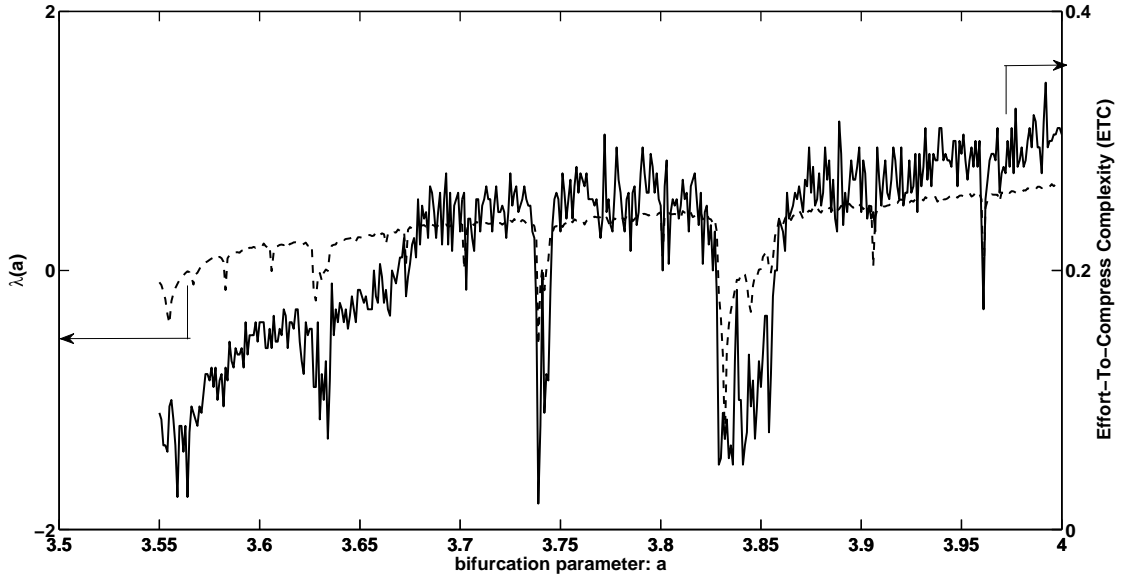}}
\caption{Effort-To-Compress complexity $ETC$ (solid) and Lyapunov exponent $\lambda(a)$ (dotted) for the logistic map as the bifurcation parameter $a$ is varied from $3.55$ to $4.0$ (in steps of $0.01$). The graphs are well correlated (positively). This is indicated by a value of Pearson's correlation coefficient $=0.8771$ between $ETC$ and $\lambda(a)$. Here, we have chosen $L=200$ and number of bins $= 4$ for the symbolic sequence.}
\label{figure:figETC}
\end{figure}

Figure~\ref{figure:figETC} depicts the graph of Effort-To-Compress complexity $ETC$ and the Lyapunov exponent $\lambda(a)$. The graphs are well correlated (positively). This is indicated by a high value of Pearson's correlation coefficient $=0.8771$ between $ETC$ and $\lambda(a)$. As the length of time series is varied, the correlation coefficient is consistently high as shown in Figure~\ref{figure:comparison} and both $LZ$ and $ETC$ outperform Shannon entropy $H$.

\subsection{CCM vs. Shannon Entropy on chaotic dynamical systems}
\label{sec:3.3} 
Inspired by the superior performance of Compression-Complexity Measures (or CCM) such as $LZ$ and $ETC$ over Shannon entropy ($H$) in characterizing the dynamical complexity of the chaotic Logistic map (as seen in the previous section), we shall now evaluate their performance on short and noisy time series from other chaotic dynamical systems - both maps and flows,  with and without noise. We consider time series from various chaotic dynamical systems for our test~\cite{Alligood}. In Table~\ref{tab:tableSystems}, we indicate the name of the chaotic system as well as the parameter settings we have chosen for our study. The goal is to evaluate whether the measures ($H, LZ$ and $ETC$) can automatically classify the time series from each system generated under different parameter settings and which is known to generate different complexities. 
\begin{table}[h]
\begin{center}
\caption{
Chaotic dynamical systems and their parameter settings chosen for the study.}
\label{tab:tableSystems}   
\begin{tabular}{|c||c|c|c|}
  \hline
  \textbf{System} & \textbf{Equation} & \textbf{Param. Settings} & \textbf{Remarks}\\
     \hline
  Logistic map    & $x_{n+1} = a x_{n}(1-x_{n})$ & $a=3.83, 3.9, 4.0$ & 1D Map \\
     \hline
  H\'{e}non   & $x_{n+1} = 1-ax_n^2+y_n$, & $a=1.2, 1.3, 1.4$, & 2D \\
  map & $y_{n+1}=bx_n$ & $b=0.3$ & Map\\
     \hline
    & $\frac{dx}{dt} = \sigma y- \sigma x$ & $\sigma = 10$, $\beta = \frac{8}{3}$, & 3D \\
Lorenz system & $\frac{dy}{dt} = \rho x - xz -y$ &  $\rho = 20, 25, 28$&  Flow\\
  & $\frac{dz}{dt} = xy + \beta z$ &  & \\
     \hline
\end{tabular}
\end{center}
\end{table}

The reason for the specific values for the choice of parameter settings is that it yields the following hierarchy of complexities (where the symbol $A$ $\prec$ $B$ means that time series corresponding to $A$ has {\it lower complexity} than that of $B$):
\begin{itemize}
    \item Logistic map:~($a=3.83$) $\prec$ ($a=3.9$) $\prec$ ($a=4.0$) $\prec$ uniform random sequence.\\
    \item H\'{e}non map: ($a=1.3$) $\prec$ ($a=1.2$) $\prec$ ($a=1.4$) $\prec$ uniform random sequence.\\
    \item Lorenz system: ($\rho=20$) $\prec$ ($\rho=25$) $\prec$ ($\rho=28$) $\prec$ uniform random sequence.\\
\end{itemize}
Note, this hierarchy is determined independently by the Lyapunov exponent in each case and serves as the gold truth for our study. Our goal is to determine whether Shannon entropy and CCMs can correctly determine this hierarchy from the given time series. Uniform random sequence has the largest Lyapunov exponent and hence it has the highest complexity.

Mehran Talbinejad \textit{et al}.~\cite{Talebinad} have done a basic complexity analysis of  data from the logistic map using the LZ complexity measure and have shown that LZ is able to distinguish between data complexities for sequences of different lengths, but the analysis is done for data generated using a single initial value. This is not enough to claim that the measures are able to distinguish data of different complexities, since different initial conditions will give rise to completely different sequences. We analyze the problem with multiple values of initial conditions and  perform statistical hypothesis testing for differences in means to determine the minimum sequence length at which correct identification is achieved. This is performed for all the three measures, namely ETC, LZ and H, and a comparative analysis is performed. This analysis is done for data generated using the one 1D map: logistic map, one 2D map:  H\'{e}non map, and one 3D flow: Lorenz system. We also include a uniform random sequence and check if the measures are able to distinguish it from chaotic data of different complexities. 

All the time series that is produced by the above mentioned chaotic systems consist of real numbers. We convert the input time series into a symbolic sequence (as described in the beginning of section 3.1). We have used four bins (which means the symbolic sequence will consist of only four symbols) in our study. Having generated different sequences of varying lengths, complexity measures are applied and the results observed to see if there are statistically significant differences in the calculated complexity values and whether the correct hierarchical ordering of the sequences based on these complexity values are obtained in each case. This is achieved by analysis using one-way ANOVA (Analysis of Variance) with post-hoc Tukey HSD (Honest Significant Difference) test for multiple comparisons. We then determine the minimum length of data needed to achieve the correct hierarchical ordering for each map and for each measure. The results are shown in Table~\ref{tab:tableNoNoise}, from which it is clear that the best (least) value of minimum length is achieved by $ETC$, followed by $LZ$. Shannon entropy's performance is the worst in each instance. Thus, CCMs outperform Shannon entropy for characterizing the dynamical complexity of short time series from chaotic maps and flows.

\begin{table}[h]
\begin{center}
\caption{Minimum length of data required for correct hierarchical ordering of sequences (from low to high complexities) for chaotic dynamical systems described in Table~\ref{tab:tableSystems}. Both $ETC$ and $LZ$ outperform Shannon entropy ($H$). $ETC$ is the best among the three measures.}
\label{tab:tableNoNoise}   
\begin{tabular}{|c||c|c|c|}
  \hline
  \textbf{Chaotic} & Min. Len. & Min Len. & Min. Len. \\
 \textbf{System} &  \textbf{H} & \textbf{ LZ} & \textbf{ETC}\\
 
     \hline
  Logistic map    & 125 & 20 & 15 \\
     \hline
  H\'{e}non map  & 1350 & 30 & 30 \\
     \hline
  Lorenz system  & $>10^4$ & 40 & 40\\
   \hline
\end{tabular}
\end{center}
\end{table}
\subsubsection{Effect of additive gaussain noise}
We investigated the effect of additive noise on the performance of the compleixty measures in characterizing the dynamical complexity of the chaotic dynamical systems. To this end, zero mean Gaussian noise with a standard deviation of 1 was added on to the time series from the chaotic dynamical systems. The symbolic sequence was extracted from the time series as before (4 bins) and similar analysis was performed to determine the minimum length of the sequence for determining the correct hierarchical ordering (from low to high complexities) in each case. To ensure that the noise doesn't override the signal itself, only a fraction of the noise output is added. In each of the cases, by trial and error, we found the signal to noise ratio (SNR) at which the performance of the measures was very close to the noise-free condition. Then for noise analysis, we considered noise with SNR that was around 15-20\% less than the SNR at which the performance matches with the noise-free condition. Table ~\ref{tab:tableWithNoise} shows the performance analysis of the different measures under noisy conditions. It is evident that all measures undergo  performance degradation due to the presence of noise. Shannon entropy undergoes more degradation than $LZ$ and $ETC$. ETC is still the best among the three measures, even in the presence of a significant amount of additive gaussian noise. We can conclude that CCMs are effective in characterizing dynamical complexity for short and noisy time series from chaotic dynamical systems.
 
\begin{table}[h]
\begin{center}
\caption{Minimum length of data required for correct hierarchical ordering of sequences (from low to high complexities)  (in the presence of additive Gaussian noise with specified $SNR$ (dB)) for chaotic dynamical systems described in Table~\ref{tab:tableSystems}. $SNR_*$ denotes the $SNR$ below which performance of the measure degrades and above which the measure is robust to noise and $SNR_{Anal}$ gives the $SNR$ used for noise analysis. $ETC$ is the best among the three measures.}
\label{tab:tableWithNoise}   
	\begin{tabular}{|c||c|c||c|c|c|}
	\hline
 \textbf{Chaotic} & \bf{ $SNR_{*}$ } & \textbf{$SNR_{Anal}$ }  & Min. Len. & Min. Len. & Min. Len. \\
 \textbf{System} & (dB) & (dB) & \textbf{H} & \textbf{ LZ} & \textbf{ETC}\\
     	\hline
  		Logistic map & 70 & 58   & 180 & 30 & 20 \\
     	\hline
  		H\'{e}non map  & 70 & 58 & $>5\times10^4$ & 40 & 40 \\
     	\hline
  		Lorenz system  & 140 & 125 & $>10^4$ & 90 & 60\\
     \hline
\end{tabular}
\end{center}
\end{table}
%
%
\begin{table}[h!]
\begin{center}
\caption{CCM vs. H on very short binary sequences. For all binary sequences (lengths $4$ to $16$), the number of distinct values for each measure as well as the mean is indicated. ETC has the highest number of distinct values which allows a better discrimination between short binary sequences.}
\label{tab:tableVeryShortBinSeq}   
	\begin{tabular}{|c||c|c|c|c|c|c|}
	\hline
\multirow{2}{*}{\textbf{Length}} & \multicolumn{2}{c|}{ \textbf{H} } & \multicolumn{2}{c|}{\textbf{ LZ} } & \multicolumn{2}{c|}{ \textbf{ ETC}}\\
     	\cline{2-7}
 	& \# distinct val.& Mean & \# distinct val. & Mean & \# distinct val. & Mean\\
     	\hline
4	&	3	&	0.7806	&	2	&	1.3750	&	3	&	0.7917 \\
\hline
5	&	3	&	0.8324	&	3	&	1.4512	&	3	&	0.8125 \\
\hline
6	&	4	&	0.8648	&	3	&	1.4540	&	5	&	0.7500 \\
\hline
7	&	4	&	0.8867	&	4	&	1.4538	&	4	&	0.7682 \\
\hline
8	&	5	&	0.9024	&	4	&	1.4590	&	6	&	0.7321 \\
\hline
9	&	5	&	0.9143	&	4	&	1.4529	&	6	&	0.7109 \\
\hline
10	&	6	&	0.9235	&	5	&	1.4456	&	8	&	0.6860 \\
\hline
11	&	6	&	0.9309	&	5	&	1.4373	&	5	&	0.6722 \\
\hline
12	&	7	&	0.9370	&	5	&	1.4304	&	10	&	0.6532 \\
\hline
13	&	7	&	0.9421	&	6	&	1.4221	&	7	&	0.6407 \\
\hline
14	&	8	&	0.9464	&	6	&	1.4137	&	9	&	0.6262 \\
\hline
15	&	8	&	0.9501	&	6	&	1.4055	&	10	&	0.6140 \\
\hline
16	&	9	&	0.9534	&	7	&	1.3980	&	11	&	0.6023 \\
     	\hline
\end{tabular}
\end{center}
\end{table}

\subsection{CCM vs. Shannon Entropy on very short binary sequences}
\label{sec:3.4} 
Looking at Table~\ref{tab:tableNoNoise}, one may wonder how do these measures (CCM and Shannon entropy) perform on very short binary sequences (length $<20$)? The motivation for investigating complexity of very short binary sequences is their occurrence in neuroscience applications where one is interested in estimating entropy/complexity of `spike trains' (membrane potential waveforms) from neurons~\cite{LZ3}. These spike trains are converted into a binary sequence by choosing a moving window and indicating whether the neuron under study fires (symbol `$1$') or not (symbol `$0$') in that window~\cite{LZ3}. In several instances, a neuron may fire only a few times ($10 - 20$ or even lesser) in the chosen window of the study. Thus, we are interested in the performance of these measures on such short binary sequences. 

To this end, we compute the entropy, LZ and ETC measures on {\it all} binary sequences of lengths varying from $4$ to $16$. In order to evaluate the performance of the measures on binary sequences, we compute the number of levels or distinct values taken up by the measure for a given length. As it can be seen from Table~\ref{tab:tableVeryShortBinSeq}, ETC has the best performance in terms of having the largest number of distinct levels for each length of the binary sequence. It is desirable to have a  large number of distinct values since this allows us to distinguish between individual binary sequences more finely. We also indicate the means for the three measures. Though we use a normalization for LZ, it yields values greater than 1 owing to the problem of finite data lengths (see~\cite{LZ1,LZ2}).  Both normalized ETC and Shannon entropy H do not have this problem and are always bounded between $0$ and $1$.

\section{Two-State Markov Chains}
\label{sec:4}
As previously noted, Markov chains are very important models in statistical mechanics as well as in several applications ranging from biomedical engineering to financial time series analysis. In this section, we are concerned with characterizing the dynamical complexity of two-state Markov chains. We first briefly introduce the notion of Markov chains and subsequently study the performance of $LZ$ and $ETC$ on two-state Markov chains.

\noindent \textit { \textbf {Definition:}} A discrete random process $X_1,X_2,...$ is said to be a first order Markov chain or a Markov process if, for all $n = 1, 2 \ldots $,
Pr($X_{n+1} = x_{n+1}|X_{n} = x_{n},X_{n-1} = x_{n-1}, \ldots, X_1 = x_1$) $=$  Pr($X_{n+1} = x_{n+1}|X_{n}$) \cite{CoverThomas},\\

where $Pr$ stands for probability, $n$ for discrete-time index and $Pr(X_{n+1} = A| X_n = B)$ stands for the conditional probability of being in state $A$ at time instant $n+1$, given that the current state is $B$ (at time instant $n$). Alternatively, we may define a first order Markov process as a random process whose future state depends only on the current state and doesn't directly depend on how the current state was reached. This makes it possible to characterize it by a probability transition matrix that defines the probability of transitions from each state to itself and to other states. The characterization of the Markov process is completed by also defining the outputs produced at each state. This leads us to the concept of a Markov information source.
A Markov information source may be thought of as a combination of a finite state Markov chain along with a function with domain the set of states S, and range, the possible outputs of all states (known as the alphabet of the source)~\cite{CoverThomas}. Thus a Markov source produces a series of outputs as it transitions from one state to another according to the state transition probabilities.

\subsection{Comparative analysis of LZ and ETC measures on a two-state Markov process}
\label{sec:4.1}

Lempel-Ziv complexity, which was introduced in section 3.1, is based on the rate of generation of new patterns in a sequence. It will be observed later that calculation of Lempel-Ziv complexity on data from a Markov process tends to reach a steady state value that is given by the entropy rate of a Markov Process. As a preview of that, we now define the entropy rate of any stochastic process.

Since a stochastic process consists of sequence of random variables, naturally we would like to know how the entropy of the sequence changes with the number of random variables $n$. This rate of growth of the entropy with $n$ is defined as the entropy rate ($H$) of a stochastic process.\\

\noindent \textit { \textbf {Definition:}} The $entropy \  rate$ of a stochastic process $\chi= \{X_i \}$ is defined by 
\begin{equation}
H(\chi) = \lim_{n \to \infty}\frac{H(X_1,X_2,.....X_n)}{n}.
\end{equation} 

\begin{figure}[!h]
\centering
\resizebox{0.5\columnwidth}{!}{
\includegraphics{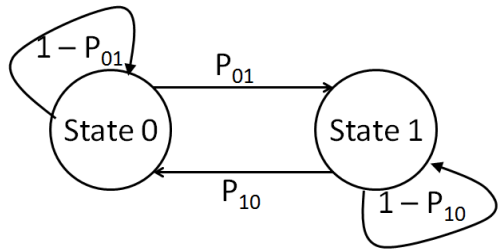}
}
\caption{A Two-state Markov Chain. The transition probabilities $P_{10}$ and $P_{01}$ correspond to transition from state 1 to 0 and 0 to 1 respectively. }
\label{figure:markov_chain}
\end{figure}

For a first order Markov chain (we assume that it is a stationary Markov process - i.e.,  state transition probabilities don't change with time) the entropy rate is defined as:
\begin{equation}
H(\chi) = \lim_{n \to \infty}H(X_n/X_{n-1}) = H(X_2/X_1) = - \sum_{ij}\mu_iP_{ij}log P_{ij}.
\end{equation}
In this equation, $\mu_i$ is the stationary probability of the $i^{th}$ state and $P_{ij}$ gives the transition probability from the $i^{th}$ state to the $j^{th}$ state. 
Specifically, for a two-state Markov chain as shown in Fig.~\ref{figure:markov_chain}, $ \mu_1 = P_{01}/(P_{01} + P_{10})$ and $\mu_2 = P_{10}/(P_{01} + P_{10}).$   \cite{CoverThomas}
 
As a first step towards the study of entropy of neural spike trains, Amigo \textit{et al.} in \cite{LZ3}, simulate a two-state Markov process as shown in Fig.~\ref{figure:markov_chain} with transition probabilities $P_{10} = 0.8$ and $P_{01} = 0.1$ and calculate normalized Lempel-Ziv complexity for varying data lengths to identify how fast it converges to a steady state value.

%
%


We recreate a similar Markov chain to generate data and find the LZ and ETC complexity measures and compare the rate at which both converge to a steady state value. Since it is a stochastic process, the simulations are run 50 times and the average value of the measures are taken. To plot the complexity values, instead of plotting values for each length, a moving window of size 20 is chosen, each window is considered as a block, and the mean values in each block is plotted. LZ complexity converges to the true entropy value~\cite{LZ3}, while the steady state value of the ETC measure is taken to be the mean ETC value in a window where the variation of the measure is less than $2\%$ of the mean value.  
\begin{figure}[!h]
\centering
\resizebox{1.0\columnwidth}{!}
{
\includegraphics{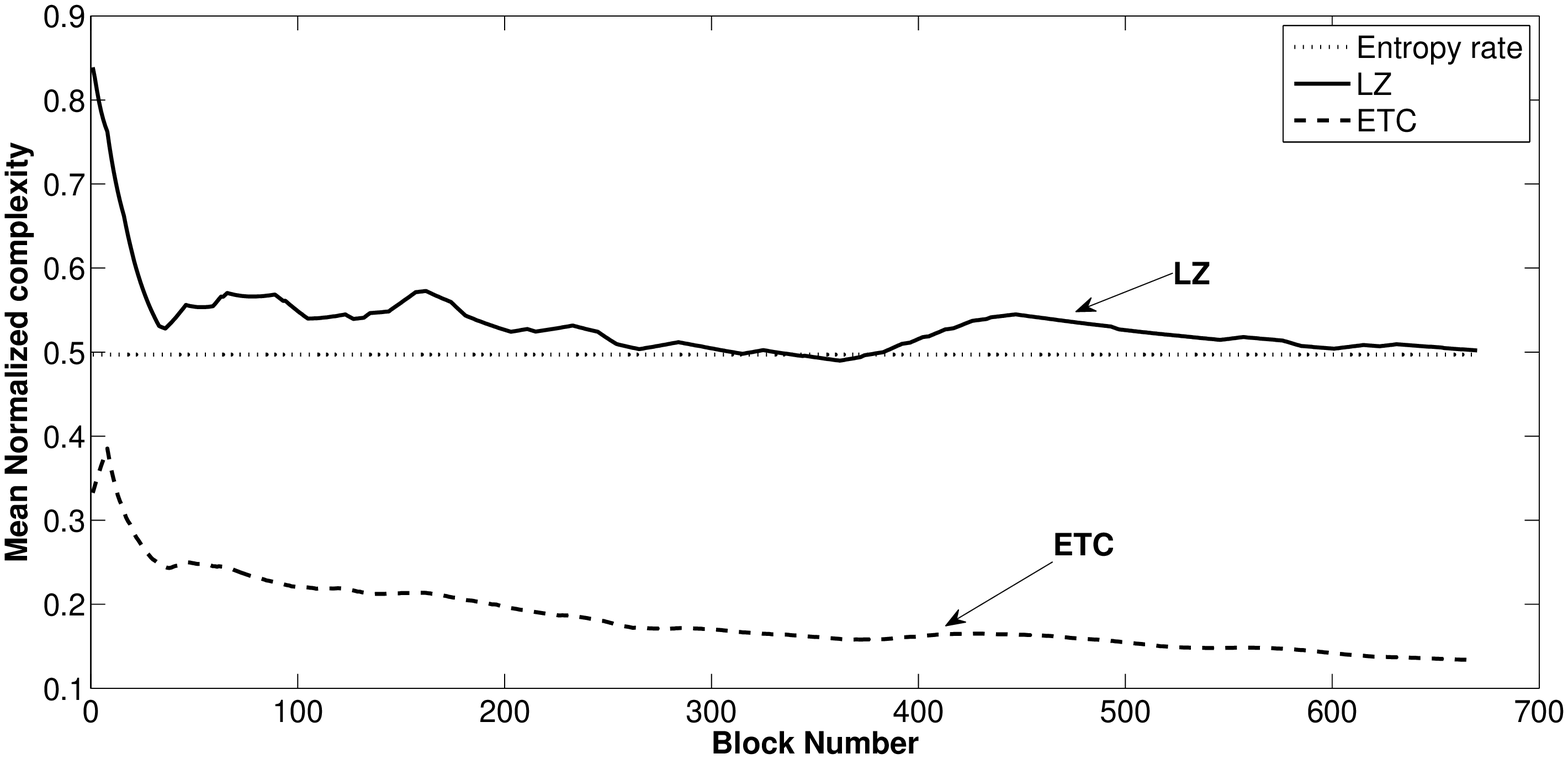}}
\caption{Convergence of ETC and LZ complexity values with increasing data length, showing faster convergence for ETC. The plot shows the average values taken over 50 iterations.}
\label{figure:ETC_markov}
\end{figure}
\begin{figure}[!h]
\centering
\resizebox{1.0\columnwidth}{!}
{
\includegraphics{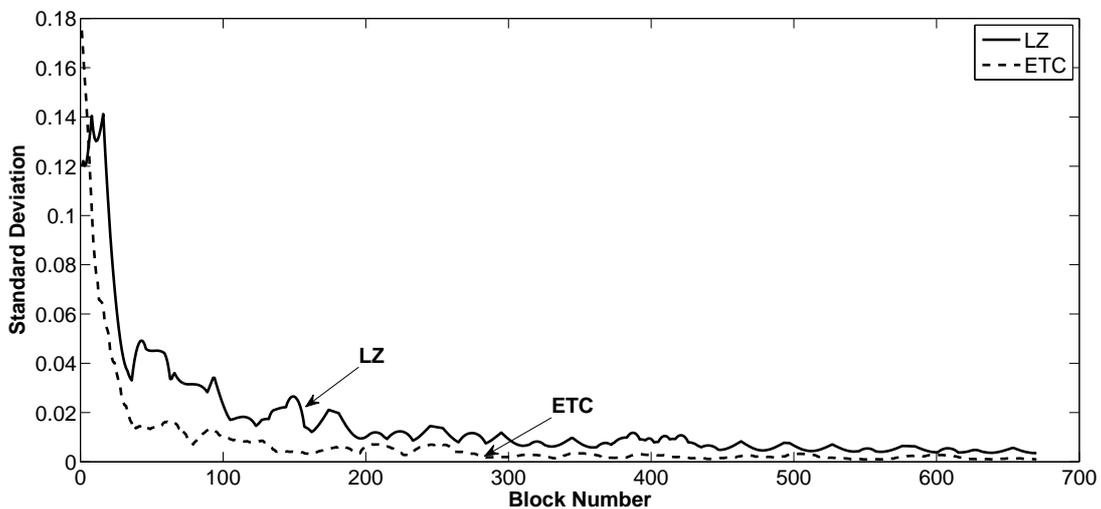}}
\caption{Standard deviation of the complexity values in each block with increasing data length.}
\label{figure:ETC_markovSTDEV}
\end{figure}
Fig.~\ref{figure:ETC_markov} shows the comparative analysis of both the measures, from which it can be seen that ETC converges faster to the steady state than LZ measure. 

Fig.~\ref{figure:ETC_markovSTDEV} shows the standard deviations of the measures in each block. It can be seen that the variations in the LZ measure are much greater than the variations in the ETC measure. This may be considered as an indicator that ETC is a more robust measure and may be used with shorter data lengths than what is possible with LZ measure. 
\section{Conclusions and Future Work}
We have considered the important problem of characterizing the dynamical complexity of short and noisy time series from chaotic dynamical systems and Markov chains, as these have practical applications in modeling. We found that Shannon Entropy is not very effective for this problem. We introduced CCMs - {\it Compression Complexity Measures} - defined as those complexity measures which are based on lossless compression algorithms. CCMs outperform Shannon entropy for characterizing complexity of both discrete and continuous chaotic dynamical systems, even in the presence of additive gaussian noise, as we have demonstrated convincingly in this work.  For 2-State Markov chains, we have empirically shown that ETC converges faster than the popular LZ complexity. Also, ETC has more number of distinct levels of complexities than H and LZ for very short binary sequences which could be potentially useful in neuroscience applications in determining complexity of spike trains. One area of future research is to determine the steady state value of ETC for Markov chains and if possible to arrive at analytical expression/bounds for the same. We also need to extend the application of ETC and LZ to Markov chains with more number of states.
%
%
%

%

\end{document}